\documentclass[aps,pre,twocolumn,superscriptaddress,tight,showpacs]{revtex4-1}

\usepackage{amssymb}
\usepackage{color}
\usepackage{graphicx}
\usepackage{amsmath}
\usepackage{mathrsfs}
\usepackage{times}
\usepackage{subeqnarray}
\usepackage{cases}
\usepackage{bm}

\begin{document}

\title{Learning the dynamics of coupled oscillators from transients}

\author{Huawei Fan}
\affiliation{School of Physics and Information Technology, Shaanxi Normal University, Xi'an 710119, China}
\affiliation{School of Science, Xi’an University of Posts and Telecommunications, Xi'an 710121, China}

\author{Liang Wang}
\affiliation{School of Physics and Information Technology, Shaanxi Normal University, Xi'an 710119, China}

\author{Yao Du}
\affiliation{School of Physics and Information Technology, Shaanxi Normal University, Xi'an 710119, China}

\author{Yafeng Wang}
\affiliation{Nonlinear Research Institute, Baoji University of Arts and Sciences, Baoji 721016, China}

\author{Jinghua Xiao}
\affiliation{School of Science, Beijing University of Posts and Telecommunications, Beijing 100876, China}

\author{Xingang Wang}
\email{Email address: wangxg@snnu.edu.cn}
\affiliation{School of Physics and Information Technology, Shaanxi Normal University, Xi'an 710119, China}

\begin{abstract}
Whereas the importance of transient dynamics to the functionality and management of complex systems has been increasingly recognized, most of the studies are based on models. Yet in realistic situations the models are often unknown and what available are only measured time series. Meanwhile, many real-world systems are dynamically stable, in the sense that the systems return to their equilibria in a short time after perturbations. This increases further the difficulty of dynamics analysis, as many information of the system dynamics are lost once the system is settled onto the equilibrium states. The question we ask is: given the transient time series of a complex dynamical system measured in the stable regime, can we infer from the data some properties of the system dynamics and make predictions, e.g., predicting the critical point where the equilibrium state becomes unstable? We show that for the typical transitions in system of coupled oscillators, including quorum sensing, amplitude death and complete synchronization, this question can be addressed by the technique of reservoir computing in machine learning. More specifically, by the transient series acquired at several states in the stable regime, we demonstrate that the trained machine is able to predict accurately not only the transient behaviors of the system in the stable regime, but also the critical point where the stable state becomes unstable. Considering the ubiquitous existence of transient activities in natural and man-made systems, the findings may have broad applications.
\end{abstract}
\date{\today }
\maketitle

\section{Introduction}

Transient activities are ubiquitous in the real world. When a neuron is stimulated by an external signal, its membrane voltage may raise temporally and then return to the resting state in milliseconds; when a pebble is thrown into a pond, ripples on the surface of the water will be disappeared in minutes; when a traffic congestion is triggered by an accident, the traffic could return to normal in hours; when an epidemic occurs, it might exist for months and finally will pass away. Roughly, transient is the temporal evolution preceding the asymptotic dynamics, which is an intrinsic property of the dynamical systems~\cite{Book:OTT,Book:Strogatz}. In traditional studies of system dynamics, attentions have been mainly focusing on the asymptotic dynamics that the system presents after the transient. Yet accumulating evidences show that transient activities might provide a new dimension in our understanding of the system dynamics~\cite{Review:Tel,Book:Lai}, and in many cases are more relevant to the system functions and performance than the asymptotic dynamics, especially for complex nonlinear systems~\cite{TRA:Hastings,TRA:Fan,TRA:Courtney1997,TRA:Mazor,TRA:MR2008,TRA:Kharin,Rev:EarthSystem2021,PowerGrid:MotterPRL2004,PowerGrid:ISPRL2008,PowerGrid:Feudel2008,Robt:KI2020,Robt:SK2020}. For instance, in ecological systems~\cite{TRA:Hastings,TRA:Fan}, transient dynamics provides a possible explanation for the occurrence of regime shift under constant environment and the sudden extinction of many species; in neuronal systems~\cite{TRA:Courtney1997,TRA:Mazor,TRA:MR2008}, neurons respond to stimulus mostly during the transients and the information encoded in the transients is more reliable than in the asymptotic dynamics; in climate systems~\cite{TRA:Kharin,Rev:EarthSystem2021}, transient dynamics provides key information for predicting the extreme events and tipping points; in power-grids~\cite{PowerGrid:MotterPRL2004,PowerGrid:ISPRL2008,PowerGrid:Feudel2008}, transient is an important concern for the system stability and security, and provides the time window for responses; in robotics~\cite{Robt:KI2020,Robt:SK2020}, transient dynamics plays a crucial role in realizing the autonomous control of robot behaviors. Whereas the importance of transient dynamics has become increasingly recognized in a wide range of disciplines, the study of transient dynamics is challenging, especially for complex systems~\cite{Review:Tel}. One reason is that the transient dynamics is different from the asymptotic dynamics. To be specific, the former is governed by the nonattracting sets embedded in the phase space, while the latter is governed by the attracting set. To analyze the transient dynamics in depth, one has to find all the nonattracting sets and their unstable manifolds~\cite{GOY:19983,KanekoChaos2003}, which are extremely difficult for high-dimensional systems~\cite{LYCPRL1995,TRA:Rempel,TRA:Morozov}. Another reason is that in realistic situations the equations governing the system dynamics are normally unknown, and what available are only the measured time series~\cite{Book:Lai}. This means that any analysis about the transient dynamics must be based on data, calling thus for the development of model-free, data-based techniques~\cite{CS:WWX2011,CS:PhyRep2016}.

Model-free prediction of chaotic systems by the technique of reservoir computer (RC) in machine learning has received considerable attention in recent years~\cite{RC:Maass2002,RC:Jaeger,RC:Lu2017,RC:Pathak2017,RC:Pathak2018,RC:Fan2020,RC:LZX2020,RC:Flynn2021-1,KLW:2021,RC:FHW2021,RC:Guo2021,RC:ZH2021,RC:Patel2021,RC:RXiao2021}. From the perspective of dynamical systems, RC can be regarded as a complex network of coupled nonlinear units which, driven by the input signals, generates the outputs through a readout function~\cite{RC:lukosevicius2009,RC:Book}. In the training phase, the input signals are provided by the target system, and the purpose of the training is to find the set of coefficients in the readout function for a best fitting of the training data. In the predicting phase, the input signals are replaced by the outputs, and the machine is running as an autonomous system with the fixed parameters. Although structurally simple, RC has shown its super power in many data-oriented applications~\cite{RC:lukosevicius2009}, e.g., speech recognization, channel equalization, robot control and chaos prediction. In particular, it has been shown that a properly trained RC is able to predict accurately the state evolution of a chaotic system for about half a dozen Lyapunov times~\cite{RC:Jaeger}, which is much longer than the prediction horizon of the traditional methods developed in nonlinear science. Besides predicting the short-term state evolution, RC is also able to replicate faithfully the long-term statistical properties of chaotic systems~\cite{RC:Pathak2017}, e.g., the dimension of strange attractors and the Lyapunov exponents. This ability, known as climate replication, has been exploited recently to predict the critical transitions in complex nonlinear systems~\cite{KLW:2021,RC:FHW2021,RC:RXiao2021}. For instance, by incorporating a parameter-control channel into RC, it is shown that the machine trained by the time series of several states in the oscillatory regime of a dynamical system is able to predict not only the critical point for system collapse, but also the averaged lifetime of the transients in the postcritical regime~\cite{KLW:2021}; training the machine by the time series of coupled oscillators at several states in the desynchronization regime, the machine is able to predict accurately the critical coupling for synchronization~\cite{RC:FHW2021}. It is noted that in predicting chaotic systems by the technique of RC, the training data are all measured from the asymptotic dynamics that the systems are finally developed to, while the transient behaviors preceding the asymptotic dynamics is normally discarded~\cite{RC:Maass2002,RC:Jaeger,RC:Lu2017,RC:Pathak2017,RC:Pathak2018,RC:Fan2020,RC:LZX2020,RC:Flynn2021-1,KLW:2021,RC:FHW2021,RC:Guo2021,RC:ZH2021,RC:Patel2021,RC:RXiao2021}.

Suppose that a complex system of coupled dynamical units is operating at a stable state (saying, for example, a steady state) and, after each perturbation, restores to the stable state after a short transient, the question we ask is: based on the transient time series measured at several states in the stable regime, can we predict the critical point where the system becomes unstable? This question is challenging and worth pursuing for several reasons. First, it is not clear whether the machine can be properly trained by the short time series measured on the transient activities. In training RC, a general requirement is that the time series should be sufficiently long (depending on the prediction tasks and the hyperparameters of the RC, the length of the training series may change from hundreds to thousands system oscillations)~\cite{RC:lukosevicius2009}. Yet in many realistic systems the transient activities sustain for only a short period (normally several system oscillations). Once settled to the asymptotic attractor, e.g., the steady state, many information of the system dynamics will be lost. The limited length of the transient dynamics therefore requires a fast learning of the machine. At present, it is not clear whether this goal can be achieved by the technique of RC. Second, different from the asymptotic dynamics, the transient dynamics is dependent on the initial conditions. Generally, the larger (smaller) is the perturbation, the longer (shorter) will be the transient~\cite{Book:OTT}. (Statistically, for the fixed system parameters, the lifetime of the transients follows an exponential distribution~\cite{Book:Lai}.) At present, it is not clear whether the machine trained by the time series of one transient process is able to predict the evolution of another transient process. Finally, even if the machine is properly trained by the transient time series, it remains unknown whether the trained machine is able to predict the critical point separating the stable and unstable regimes, as the system dynamics in the two regimes are distinctly different. Our main objective in the present work is to provide an affirmative answer to the above question. Specifically, we are going to demonstrate that for the typical bifurcation scenarios in coupled oscillators, including quorum sensing~\cite{QUO:Taylor}, amplitude death~\cite{AD:1990}, and complete synchronization~\cite{SYNREV:Pecora}, the machine trained by the transient time series acquired in the stable regime is able to predict not only the system evolution in the stable regime, but also the critical point where the stable state becomes unstable.

The rest of the paper is organized as follows. In Sec. II, we will propose the scheme of parameter-aware RC, and present the details of the training and predicting phases. The results predicted by the machine on the phase transition of three typical models, including an ensemble of indirectly coupled limit-cycle oscillators showing the phenomenon of quorum sensing, two coupled non-identical limit-cycle oscillators showing the phenomenon of amplitude death, and two coupled chaotic oscillators showing the phenomenon of complete synchronization, will be presented in Sec. III. Discussions and conclusion will be given in Sec. IV.   

\section{Parameter-aware reservoir computer}\label{RC}

We generalize the scheme of parameter-aware RC proposed in Refs.~\cite{KLW:2021,RC:FHW2021} to predict the phase transition in coupled oscillators. The RC consists of four components: the $I/R$ layer (input-to-reservoir), the parameter-control channel, the reservoir, and the $R/O$ layer (reservoir-to-output). The $I/R$ layer is characterized by the matrix $\mathbf{W}_{in}\in\mathbb{R}^{D_r\times D_{in}}$, which couples the input vector $\mathbf{u}_{\beta}(t)\in\mathbb{R}^{D_{in}}$ to the reservoir network. Here, $\mathbf{u}_{\beta}(t)$ denotes the input vector that is acquired from the target system at time $t$ and under the specific bifurcation parameter $\beta$. The elements of $\mathbf {W}_{in}$ are randomly drawn from a uniform distribution within the range $[-\sigma, \sigma]$. The parameter-control channel is characterized by the vector $\mathbf{s}=\beta\mathbf{b}$, with $\beta$ the control parameter and $\mathbf{b}\in \mathbb{R}^{D_{r}}$ the bias vector. In applications, the control parameter $\beta$ can be treated as an additional input channel marking the input vector $\mathbf{u}(t)$. The elements of $\mathbf{b}$ are drawn randomly from a uniform distribution within the range $[-\sigma, \sigma]$. The reservoir network contains $D_r$ dynamical nodes, with the initial states of the nodes being randomly chosen from the interval $[-1,1]$. The states of the nodes in the reservoir network, $\mathbf{r}(t)\in \mathbb{R}^{D_r}$, are updated according to the equation
\begin{equation}\label{rc1}
\mathbf{r}(t+\Delta t)=\tanh[\mathbf {A}\mathbf{r}(t)+\mathbf{W}_{in}\mathbf{u}_{\beta}(t)+\beta\mathbf{b}].
\end{equation}
Here, $\Delta t$ is the time step for updating the reservoir network, $\mathbf{A}\in \mathbb{R}^{D_r\times D_r}$ is the weighted adjacency matrix representing the coupling relationship between nodes in the reservoir. The adjacency matrix $\mathbf{A}$ is constructed as a sparse random Erd\"{o}s-R\'{e}nyi matrix: with the probability $p$, each element of the matrix is arranged a nonzero value drawn randomly from the interval $[-1,1]$. The matrix $\mathbf{A}$ is rescaled to make its spectral radius equal $\lambda$. Before the training process, the reservoir is evolved for a transient period of $T'_0$, so as to avoid the influence induced by the initial states of the nodes. The output layer is characterized by the matrix $\mathbf{W}_{out}\in \mathbb{R}^{D_{out}\times D_{r}}$, which generates the output vector, $\mathbf{v}(t)\in \mathbb{R}^{D_{out}}$, according to the equation 
\begin{equation}\label{rc2}
\mathbf{v}(t+\Delta t)=\mathbf{W}_{out}\mathbf{\tilde{r}}(t+\Delta t),
\end{equation}
with $\mathbf{\tilde{r}}\in \mathbb{R}^{D_r}$ the new sate vector transformed from the reservoir state (i.e., $\tilde{r}_i=r_i$ for the odd nodes and $\tilde{r}_i=r_i^2$ for the even nodes)~\cite{RC:Pathak2018}, and $\mathbf{W}_{out}$ the output matrix to be obtained by the training process. Except $\mathbf{W}_{out}$, all other parameters of the RC, e.g., $\mathbf{W}_{in}$, $\mathbf{A}$ and $\mathbf{b}$, are fixed at the construction. The purpose of the training process is to find a suitable output matrix $\mathbf{W}_{out}$ so that the output vector $\mathbf{v}(t+\Delta t)$ as calculated by Eq. (\ref{rc2}) is as close as possible to the input vector $\mathbf{u}(t+\Delta t)$ for $t=(\tau+1)\Delta t,\ldots,(\tau+L)\Delta t$, with $T_0=\tau\Delta t$ the transient period and $L$ the length of the training time series. This can be done by minimizing the following the cost function with respect to $\mathbf{W}_{out}$~\cite{RC:Lu2017,RC:Pathak2017,RC:Pathak2018}
\begin{equation}\label{rc3}
\mathbf{W}_{out}=\mathbf{U}\mathbf{V}^T(\mathbf{V}\mathbf{V}^T+\eta\mathbb{I})^{-1}.
\end{equation}
Here, $\mathbf{V}\in \mathbb{R}^{D_{r}\times L}$ is the state matrix whose $k$th column is $\mathbf{\tilde{r}}[(\tau+k)\Delta t]$, $\mathbf{U}\in \mathbb{R}^{D_{in}\times L}$ is a matrix whose $k$th column is $\mathbf{u}[(\tau+k)\Delta t]$, $\mathbb{I}$ is the identity matrix, and $\eta$ is the ridge regression parameter for avoiding the overfitting. After training, the output matrix $\mathbf{W}_{out}$ will be fixed, and the RC is ready for prediction. In the predicting phase, first we set the control parameter $\beta$ to a specific value of interest (not necessarily the parameters used in the training phase), then we evolve the RC as an autonomous dynamical system by taking the output vector $\mathbf{v}(t)$ as the input vector at the next time step $\mathbf{u}_{\beta}(t+\Delta t)$. The output vector $\mathbf{v}(t)$ gives the prediction. For the sake of simplicity, we set $D_{out}=D_{in}$~\cite{RC:Lu2017,RC:Pathak2017,RC:Pathak2018}.

We note that in the training phase the input data consists of two time series: (1) the input vector $\mathbf{u}_{\beta}(t)$ representing the state of the target system and (2) the control parameter $\beta(t)$ labeling the condition under which the input vector $\mathbf{u}_{\beta}(t)$ is acquired. In specific, the input vector $\mathbf{u}_{\beta}(t)$ is composed of $m$ segments of length $L'$, while each segment is a time series obtained from the target system under the specific control parameter $\beta$. As such, $\beta(t)$ is a step-function of time. In the predicting phase, besides replacing $\mathbf{u}_{\beta}(t)$ with $\mathbf{v}(t)$, we still need to input the control parameter $\beta(t)$ in a step-wise fashion, so as to guide the reservoir evolution. The essential difference between the current study and the previous ones lies in the regime where the training data are acquired. In Ref.~\cite{KLW:2021}, the training data are taken from the oscillatory regime of chaotic attractors, and the mission is to predict the critical point leading to the steady states; in Ref.~\cite{RC:FHW2021}, the training data are acquired from the desynchronization regime, and the mission is to predict the critical coupling generating complete synchronization. In both studies, the measured time series are sufficiently long for training the machine. Different from these studies, in our present work the training data are taken from the regime of steady states (or complete synchronization state), in which only a short time series, i.e., the transient dynamics, are available for training the machine. 

\section{Results}\label{app}

\subsection{Quorum sensing in coupled Stuart-Landau oscillators}\label{qs}

Quorum sensing refers to the sudden switch of the dynamics of an ensemble elements from the quiescent state to the synchronized oscillatory state as the population density of the elements exceeds a critical value, which is normally observed in systems where a large population of dynamical elements are coupled indirectly through a common medium, e.g., chemical oscillators~\cite{QUO:Taylor}, bacterial systems~\cite{QUO:MBM2001}, genetic cellular oscillators~\cite{QUO:Garcia}, and crowd on a footbridge~\cite{QUO:Stogatz2005}. Below the critical population density, the quiescent state is stable, i.e., once perturbed, the system will return to the quiescent state after a short transient. Assuming that the system is operating in the quiescent regime and only the transient series at several population densities are available, the question we ask is, without knowing the equations of the system dynamics, can we predict when will the elements start to oscillate? We are going to demonstrate that for the typical model showing quorum sensing, the question can be addressed by the scheme of parameter-aware RC introduced in Sec.~\ref{RC}.

The model for quorum sensing we adopt here is an ensemble of Stuart-Landau oscillators coupled through an external medium~\cite{QUO:Schwab}. The system dynamics reads
\begin{align}
&\dot{z}_{j}=(1 + i\omega_{j} - |z_{j}|^{2})z_{j}+ D(Z-z_{j}), \\
&\dot{Z}=-(J+i\omega_{0})Z+\frac{\rho D}{N}\sum^{N}_{j=1}(z_{j}-Z),
\end{align}
where $j=1,2,...,N$ is the oscillator index, $z_{j}=x_{j}+iy_{j}$ is the complex variable denoting the state of oscillator $j$, and $Z=X+iY$ is the complex variable denoting the state of the external medium. In Eq. (4), $\omega_{j}$ is the natural frequency of oscillator $j$, which is randomly chosen within the range $[-1,1]$. The oscillators are isolated from each other, but are all coupled to the external medium through the diffusions, with the coupling strength (diffusing coefficient) being represented by $D$. In Eq. (5), $\omega_0$ denotes the natural frequency of the medium, $J$ represents the relaxation parameter, and $\rho$ denotes the population density, which plays as the bifurcation parameter for generating quorum sensing. (As the number of oscillators is fixed, the increase of the population density corresponds to the decrease of the system volume in experiments~\cite{QUO:Schwab,QUO:Taylor,QUO:MBM2001,QUO:Garcia,QUO:Stogatz2005,QUO:Li}.) In simulations, we set the system parameters as $N=100$, $D=1.5$, $\omega_{0}=1$, and $J=0.2$. The initial states of the oscillators and the external medium (including the real and imaginary parts) are randomly chosen from the range $(-\delta, \delta)$. The equations are solved numerically by the fourth-order Runge-Kutta method with the time step $\Delta t=0.1$. As in Refs.~\cite{QUO:Schwab,QUO:Taylor,QUO:MBM2001,QUO:Garcia,QUO:Stogatz2005,QUO:Li}, we characterize the asymptotic behaviors of the oscillators and the external medium by the the order parameters $R_{z}=\langle|\frac{1}{N}\sum^{N}_{j=1}z_{j}|\rangle_{T}$ and $R=\langle|Z|\rangle_{T}$, respectively. Here, $\langle...\rangle_{T}$ denotes that the results are averaged over a time period of $T$. To focus on the asymptotic behaviors of the system, a transient period of $T'$ has been discarded in calculating the order parameters. 

\begin{figure*}[tpb]
\begin{center}
\includegraphics[width=0.95\linewidth]{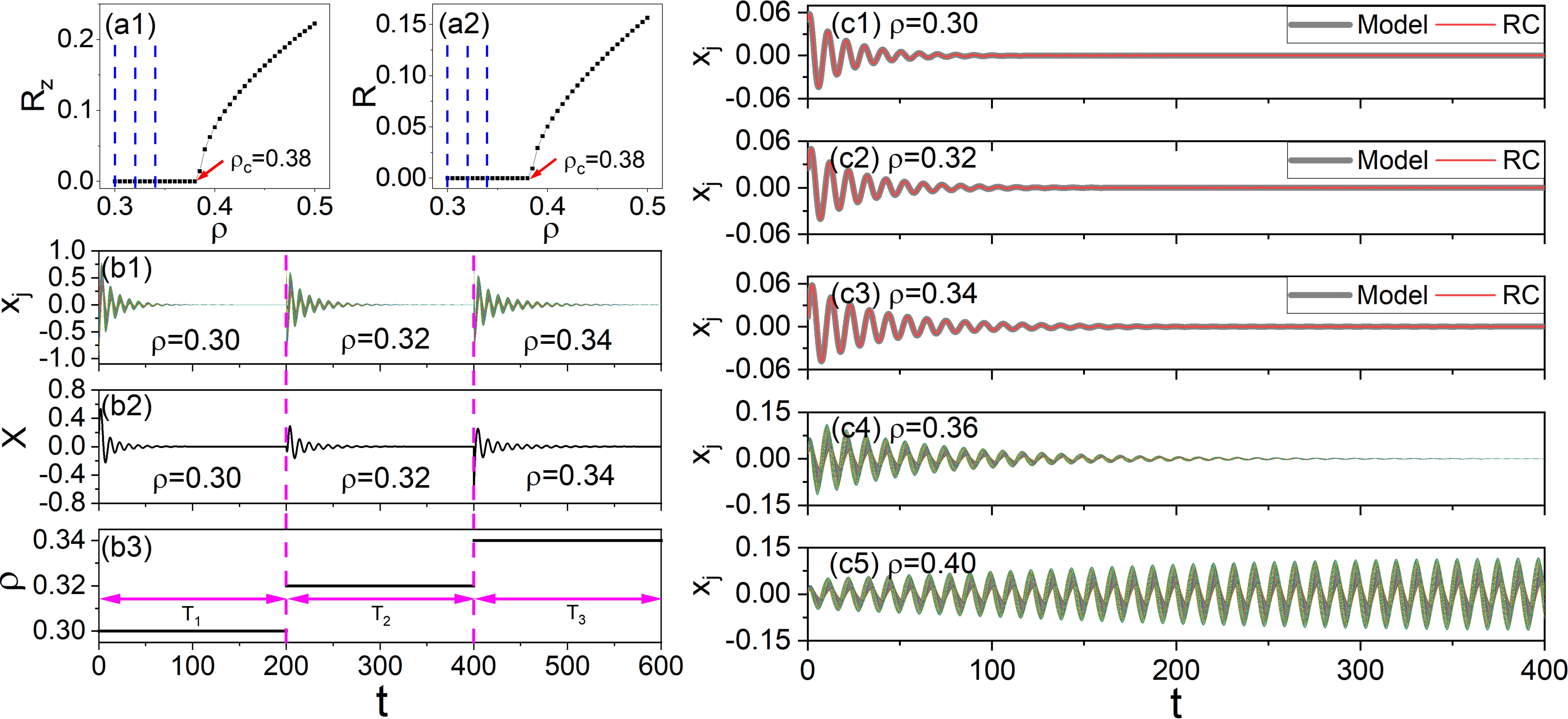}
\caption{Predicting the dynamics of an ensemble of Stuart-Landau oscillators coupled through an external medium. (a1,a2) The variations of the system order parameters $R_z$ and $R$ with respect to the population density $\rho$ obtained from model simulations. The transition from the quiescent to oscillatory states occurs at $\rho_c\approx 0.38$. Vertical lines denote the sampling states where the training data are acquired. (b1,b2) The time evolution of the oscillators and the medium under different values of $\rho$ adopted in the quiescent regime. (b3) The time series of the control parameter. $T_1=T_2=T_3=200$ is the time length of the sampling time series. (c1-c5) The time evolutions of the system predicted by the machine for different values of $\rho$. Black curves are the results obtained from model simulations. Red curves are the results predicted by the machine.}
\label{fig1}
\end{center}
\end{figure*}

Setting $\delta=1$, $T'=1\times 10^4$ and $T=2\times 10^4$, we plot in Fig.~\ref{fig1}(a) the variations of $R_z$ and $R$ with respect to $\rho$. It is seen that $R_z=R=0$ for $\rho<\rho_c\approx 0.38$ and, as $\rho$ exceeds $\rho_c$, the values of $R_z$ and $R$ are gradually increased. That is, the asymptotic dynamics of the oscillators and the medium is quiescent for $\rho<\rho_c$ and is oscillatory for $\rho>\rho_c$. To show the transient dynamics of the system in the quiescent regime, we plot in Figs.~\ref{fig1}(b1) and (b2) the time evolution of the oscillators and the medium for $m=3$ sampling population densities ($\rho=0.30$, $0.32$ and $0.34$), each of time duration $T=200$. It is seen that in all three cases, the oscillators and the medium are damped to the origin after a short transient ($t< 100$). Here our mission is to infer from the transient time series measured at the three sampling states the transition point $\rho_c$.  

To implement the parameter-aware RC, we first generate the training data by combining the time series of the three sampling states into a long time sequence. The training series are plotted in Figs.~\ref{fig1}(b1) and (b2). As the time series of each sampling state contains $L'=T/\Delta t = 2\times 10^3$ data points, the length of the combined sequence therefore is $L=mL'=6\times 10^3$. The corresponding time series of the population density, as shown in Fig.~\ref{fig1}(b3), is used as the inputs of the parameter-control channel, i.e., replacing $\beta$ with $\rho$ in Eq.~(\ref{rc1}). Then, using the combined series and $\rho(t)$ as the inputs, we calculate the output matrix according to Eq.~(\ref{rc3}). This completes the training phase. 

In obtaining the output matrix, a short segment of $\tau=100$ data points in each sampling series is used to drive the reservoir out of the transient. The hyperparameters parameters of RC are chosen as $(D_{r}, p, \sigma, \lambda, \eta)=(5\times 10^3, 0.1, 1, 0.1, 1\times 10^{-9})$, which are obtained by the optimizer ``optimoptions" in MATLAB. In addition, to reduce the complexity of the input matrix $\mathbf{W}_{in}$, we set only one non-zero element in each row of the input matrix. That is, each node in the reservoir network receives only one component from the input vector $\bm{u}=[x_{1},y_{1},...,x_{N},y_{N},X,Y]^{T}$.

Before predicting the transition point of quorum sensing, we check first the performance of the trained machine in predicting the transient dynamics of the sampling parameters, including $\rho=0.30$, $0.32$ and $0.34$. In the predicting phase, we replace the input vector $\mathbf{u}$ in Eq.~(\ref{rc1}) with the output vector $\mathbf{v}$ calculated by Eq.~(\ref{rc2}), while setting the control parameter $\beta$ in Eq.~(\ref{rc1}) to be one of the sampling parameters. The reservoir network is started from the random initial conditions, and is driven by the testing data for $\tau'=10$ time steps~\cite{RC:Jiang}. (This strategy of ``cold start" is necessary for predicting the time evolution of the system state, but is not when the mission is to predict the transition point~\cite{KLW:2021,RC:FHW2021,RC:ZH2021}.) The predicted results are plotted in Figs.~\ref{fig1}(c1-c3). We see that in all the cases, the machine predicts accurately not only the transient evolution of the system state, but also the quiescent state the system is finally settled to.

\begin{figure*}[tbp]
\begin{center}
\includegraphics[width=0.8\linewidth]{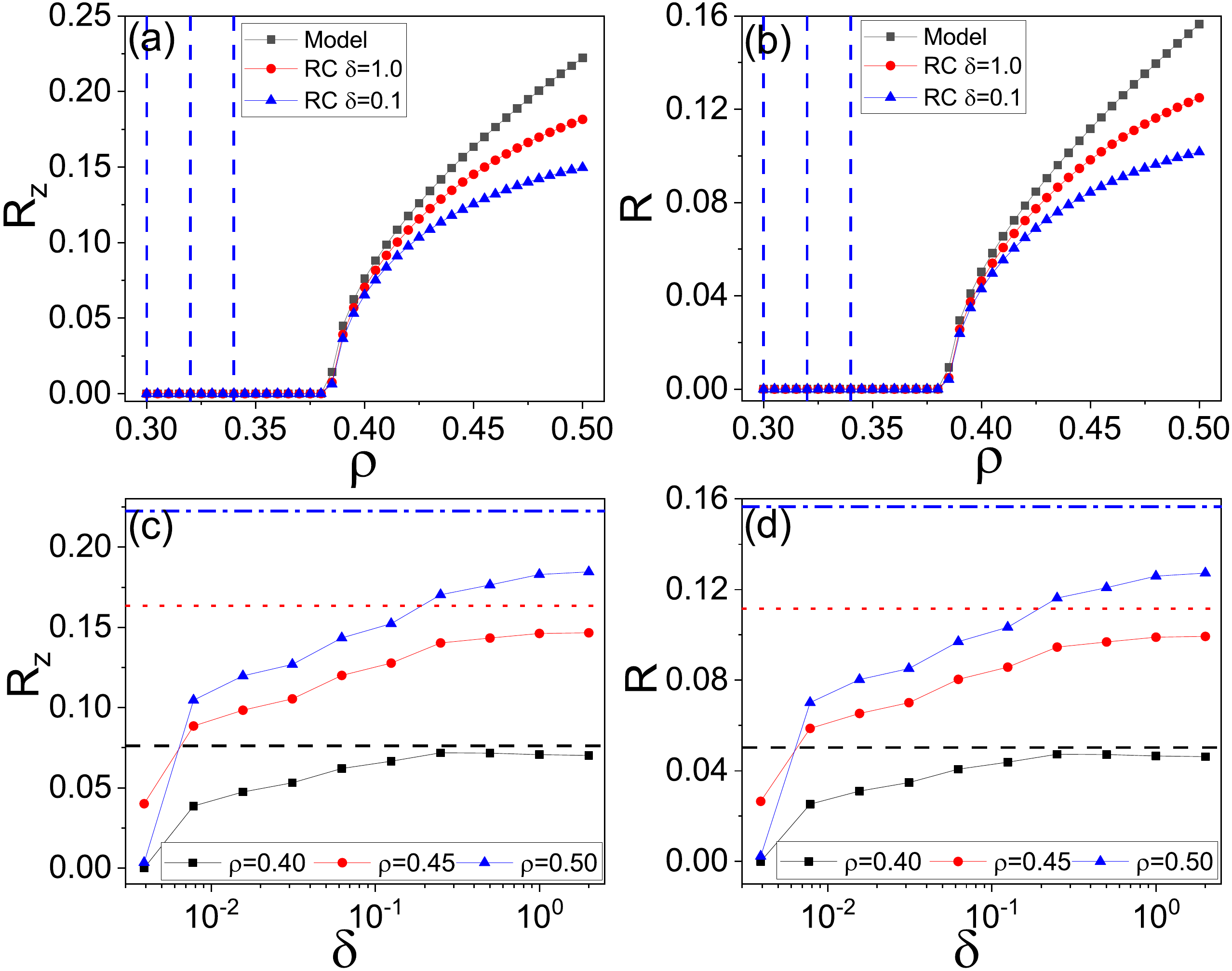}
\caption{Predicting quorum-sensing transition in an ensemble of indirectly coupled Stuart-Landau oscillators. (a,b) The variation of the order parameters $R_z$ and $R$ with respect to $\rho$ predicted by machine. The initial conditions of the system are randomly chosen from the range $(-\delta,\delta)$. Black squares: the results obtained by model simulations. Red discs: the predicted results for $\delta=1.0$. Blue triangles: the predicted results for $\delta=0.1$. Vertical lines denote the sampling states. (c,d) The variation of $R_z$ and $R$ with respect to $\delta$ for different values of $\rho$ adopted in the oscillatory regime. Horizontal lines are the corresponding order parameters obtained from model simulations. As $\delta$ increases, the predicted order parameters approach the actual values.}
\label{fig2}
\end{center}
\end{figure*}

We check further the capability of the trained machine in predicting the time evolution of the system state under a new parameter that is not included in the sampling set. To demonstrate, we set $\rho=0.36$ as the control parameter, which is also within the quiescent regime. The predicted results for this new parameter are plotted in Fig.~\ref{fig1}(c4). It is seen that the transient dynamics and the quiescent state are well predicted by the machine too. Setting $\rho=0.4$, we plot in Fig.~\ref{fig1}(c5) the time evolution predicted by the machine, which suggests that, instead of reaching the quiescent state, the system will finally settle to an oscillatory state. That is, the machine predicts that quorum sensing occurs before the population density $\rho=0.4$. This prediction is consistent with the results obtained from model simulations, as depicted in Fig.~\ref{fig1}(a). 

Having justified the capability of RC in predicting the dynamics of the individual states in both the quiescent and oscillatory regimes, we proceed to predict the transition point $\rho_{c}$ by the the scheme of parameter-aware RC. In doing this, we increase the control parameter $\rho$ gradually from $0.3$ to $0.5$ by the step $\Delta\rho=0.05$. For each value of $\rho$, we first run the machine (starting from the random initial conditions) for a transient period of $T'_0=1\times 10^3$, and then calculate from the outputs the time-averaged order parameters $R_{z}$ and $R$ over a period of $T=1\times 10^4$. The results are plotted in Figs.~\ref{fig2}(a) and (b) (red discs). We see that the machine predicts accurately the critical population density, $\rho_c\approx 0.38$, and also the progressive increase of $R_z$ and $R$ as $\rho$ increases from $\rho_c$. Figures~\ref{fig2}(a) and (b) also show that in the oscillatory regime ($\rho>\rho_c$), the values of $R_z$ and $R$ predicted by the machine are smaller to the values obtained from model simulations. More specifically, as $\rho$ increases from $\rho_c$, the error between the predicted and actual results is gradually enlarged. 

As the training data are collected from states in the quiescent regime, the useful information for machine training is thus contained in the transient activities only. It is natural to conjecture that the longer the transient, the more information will the machine infer from the data, and the more accurate will be the prediction. To justify this conjecture, we enlarge the range $(-\delta,\delta)$ over which the initial conditions of the oscillators and the medium are chosen, and check again the prediction performance by training a new machine. Statistically, with the increase (decrease) of $\delta$, the lifetime of the transient will be extended (shortened)~\cite{Book:Lai}.) Decreasing $\delta$ to $0.1$, we generate the training data by the same sampling states in the quiescent regime ($\rho=0.30$, $0.32$ and $0.34$), and train the RC again. The length of the sampling series and the hyperparameters of the RC are the same to the one used in Fig.~\ref{fig1}. The results predicted by the new machine are plotted in Figs.~\ref{fig2}(a) and (b) (blue triangles). We see that the transition point is also accurately predicted, but, comparing with the results of $\delta=1.0$, the prediction is worse in the oscillatory regime. To explore further the impact of $\delta$ on the prediction performance, we plot in Figs.~\ref{fig2}(c) and (d) the variation of the order parameters with respect to $\delta$ for different values of $\rho$ in the oscillatory regime. It is seen that with the increase of $\delta$, the order parameters predicted by the machine approach gradually the actual order parameters calculated from model simulations. Numerical evidences thus suggest that it is the transient activities that provide the useful information for machine training, and longer transient gives better predictions.   

Additional simulations have been conducted to check the impacts of the sampling states on the prediction performance, including the number of the sampling states and the locations of the sampling states. The general findings are the following (not shown): (1) when the number of sampling parameters is fixed, the closer the sampling parameters to the transition point, the better is the prediction; (2) the prediction performance is improved by adopting more sampling states in the quiescent regime. The additional results are understandable, as: (1) in the quorum-sensing transition, as the bifurcation parameter approaches the transition point from below, the lifetime of the transient will be gradually increased~\cite{QUO:Schwab,QUO:MBM2001,QUO:Garcia,QUO:Stogatz2005,QUO:Li}; (2) by combining the time series of more sampling states, the total number of transient data in the training series will be increased. These additional results suggest again that the useful information for training the machine comes from the transient dynamics, but not the asymptotic dynamics associated with the quiescent state. 

\subsection{Amplitude death in coupled Stuart-Landau oscillators}\label{ad}

Amplitude death refers to the cessation of oscillation in coupled oscillators as the system bifurcation parameter passes through a critical value~\cite{AD:1990,AD:DGA1990}. For its important implications to the functioning of many real-world systems, the phenomenon of amplitude death has been extensively studied by researchers in different fields in the past decades~\cite{AD:ZWPhyRep2021}. Similar to the phenomenon of quorum sensing, in amplitude death the system dynamics is also transited from the quiescent to oscillatory states at a critical point. To generate amplitude death, a general approach is introducing a parameter mismatch among the oscillators and using the coupling strength of the oscillators as the bifurcation parameter~\cite{AD:Timedelay}. This phenomenon can also be observed in systems of identical oscillators by adopting new coupling schemes, such as introducing time-delay to the couplings~\cite{AD:Timedelay}, using the conjugate coupling functions or dynamical couplings~\cite{AD:Conjugate,AD:Dynamical}. Recently, stimulated by the blooming of network science, amplitude death in networked oscillators has been explored, in which the important role of network structure on the transition has been revealed~\cite{AD:Hou,AD:YJZ,AD:LWQ2009}. Depending on the system functions, amplitude death might be desired or undesired. In neuronal systems, an ensemble of neural cells may stop pulsating under strong couplings, leading to the pathological conditions~\cite{AD:cell}. In physiological context, the quenching of oscillations means the loss of rhythms, resulting in diseases such as sudden cardiac death~\cite{AD:heart}. In cases like this, amplitude death is undesired and the mission is to prevent its occurrence. However, in physical systems such as coupled lasers~\cite{AD:laser}, quenched oscillations are necessary for generating a stable output; in ecological systems, large-amplitude oscillations increase the risk of species extinctions~\cite{AD:ecology}. In cases like this, amplitude death is desired and the mission becomes maintaining the quiescent state.

As in realistic situations the system equations are generally unknown, the prediction of amplitude death therefore relies on the development of model-free techniques. By the scheme of parameter-aware RC, in Ref.~\cite{RC:RXiao2021} the authors is able to predict successfully the transition to amplitude death in coupled chaotic and periodic oscillators. Different from our present work, in Ref.~\cite{RC:RXiao2021} the training data are collected in the oscillatory regime, in which the length of the measured time series at each sampling state is sufficiently long (about a few hundreds oscillations.) The question we ask is: if the time series are collected in the quiescent regime in which the transient activities sustain for only a short period (e.g., a few dozens oscillations), can we predict the critical point where the systems is survived from amplitude death? We are going to show next that for the model of coupled limit cycles, the prediction is feasible.

\begin{figure}[tbp]
\begin{center}
\includegraphics[width=0.95\linewidth]{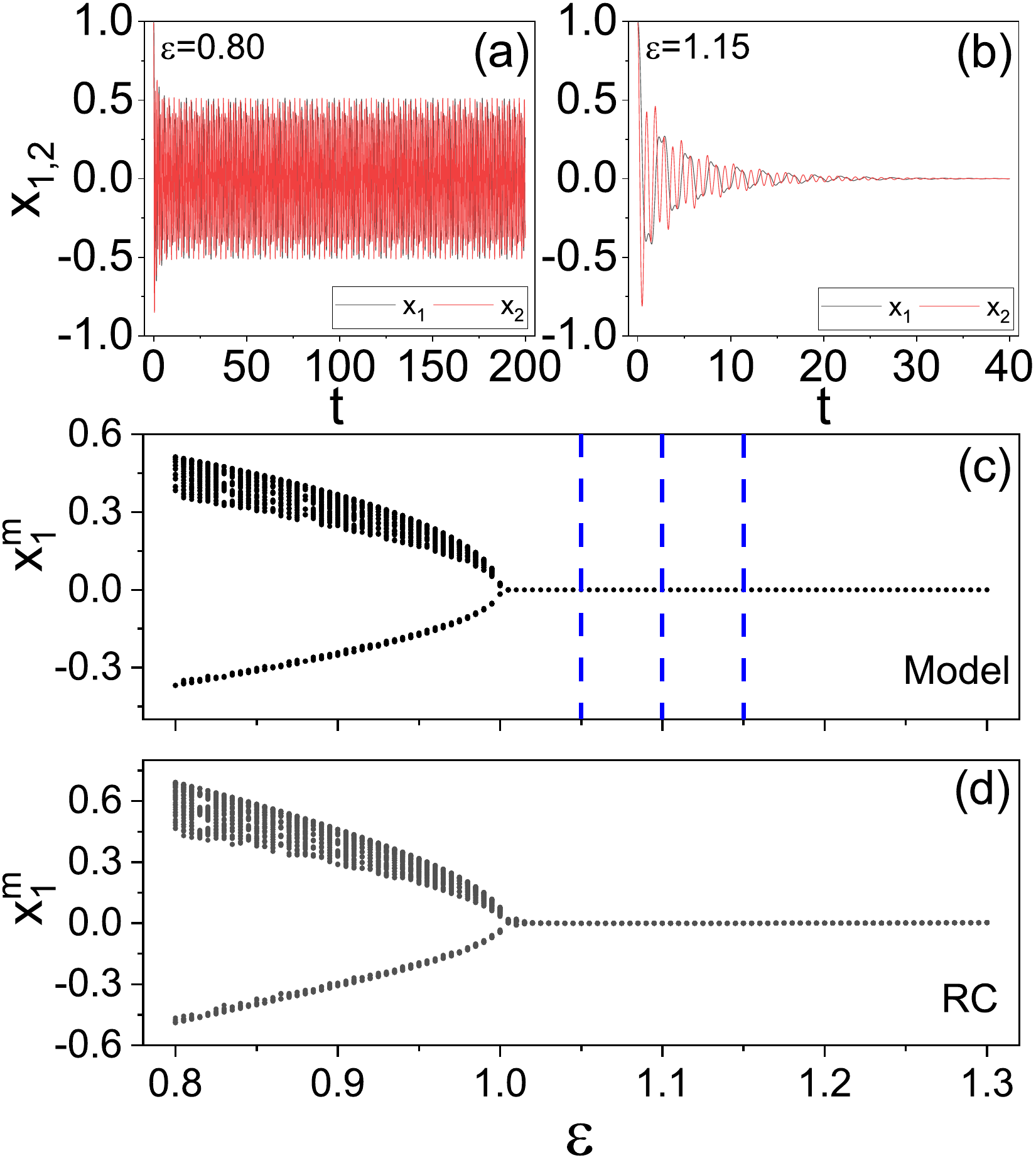}
\caption{Predicting the transition from oscillation to amplitude death in two coupled non-identical Stuart-Landau oscillators. (a) The time evolution of $x_1$ and $x_2$ for the coupling strength $\varepsilon=0.80$. (b) The time evolution of $x_1$ and $x_2$ for the coupling strength $\varepsilon=1.15$. (c) The bifurcation diagram obtained from model simulations. $x_1^m$ is the local maximum or minimum of the variable $x_1$ during the course of system evolution. Vertical lines denote the locations of the sampling states. (d) The bifurcation diagram predicted by the machine.}
\label{fig3}
\end{center}
\end{figure}

The model of amplitude death we consider here consists of two diffusively coupled non-identical Stuart-Landau oscillators~\cite{AD:1990,AD:DGA1990}, 
\begin{align}\label{ad}
&\dot{z}_{1}=(1 + i\omega_{1} - |z_{1}|^{2})z_{1}+ \varepsilon(z_{2}-z_{1}), \notag\\
&\dot{z}_{2}=(1 + i\omega_{2} - |z_{2}|^{2})z_{2}+ \varepsilon(z_{1}-z_{2}),  
\end{align}
where $z_{1,2}=x_{1,2}+iy_{1,2}$ are the complex variables denoting the oscillator states, and $\varepsilon$ is the coupling strength. As analyzed in Ref.~\cite{AD:DGA1990}, the two oscillators are ceased to the origin ($z_1=z_2=0$) when $\varepsilon>\varepsilon_c=1$ and $\Delta \omega>2\sqrt{2\varepsilon-1}$, with $\Delta \omega\equiv |\omega_1-\omega_2|$ the frequency mismatch between the oscillators. In simulations, we fix the natural frequencies of the two oscillators as $\omega_1=2.0$ and $\omega_2=7.0$, and tune the coupling strength $\varepsilon$ to investigate the transition. The initial states of the oscillators, i.e., the values of $x_{1,2}$ and $y_{1,2}$ at $t=0$, are chosen randomly from the range $(-1,1)$. The time step used in simulating Eqs. (6) is $\Delta t=0.02$. Setting $\varepsilon=0.8<\varepsilon_c$, we plot in Fig.~\ref{fig3}(a) the evolution of $x_1$ and $x_2$ with time. We see that the system shows the feature of permanent oscillation. An example of the quenched state is plotted in Fig.~\ref{fig3}(b), in which the coupling strength is set as $\varepsilon=1.15>\varepsilon_c$. We see that in this case the oscillation is ceased after a transient period about $T=20$. To have a global picture on the bifurcation diagram, we plot in Fig.~\ref{fig3}(c) the variation of the local extrema of $x_1$ with respect to $\varepsilon$, which confirms the theoretical prediction that amplitude death occurs at $\varepsilon_c=1$. Here our mission is to reproduce the bifurcation diagram based on the transient time series acquired at several states in the death regime.

In predicting the bifurcation diagram, we first generate the training data by combining the time series of several states in the death regime. As illustration, we choose the sampling states at three coupling parameters: $\varepsilon=1.05$, $1.1$, and $1.15$. The length of the time series for each state is $T=40$, and the training data contains in total $L=mT/\Delta t=6\times 10^3$ points. We then input the training data and the corresponding time series of the control parameter, $\varepsilon(t)$, to the machine, and estimate the output matrix according to Eq.~(\ref{rc3}). In this application, the optimized hyperparameters of the RC are $(D_{r}, p, \sigma, \lambda, \eta)=(600, 0.1, 1, 0.1, 1\times 10^{-7})$. Finally, we tune the control parameter $\varepsilon$ to different values, and calculate from the machine outputs the variation of the system dynamics with respect to $\varepsilon$. The predicted results are plotted in Fig.~\ref{fig3}(d), in which $x_1^m$ denotes the local maximum or minimum of the variable $x_1$ during the course of system evolution. Comparing with the results of model simulation shown in Fig.~\ref{fig3}(c), we see that both the location of the transition point $\varepsilon_c$ and the oscillating amplitudes in the oscillatory regime are well predicted by the machine.

\subsection{Complete synchronization in coupled chaotic oscillators}\label{syn}

We finally employ the scheme of parameter-aware RC to predict the transition to complete synchronization in coupled chaotic oscillators. As a universal concept in nonlinear science, synchronization has been extensively studied in literature for decades~\cite{Syn:Book}. Briefly, synchronization refers to the coherent motion of coupled oscillators, which occurs normally when the interacting strength of the oscillators is larger to a critical value. Depending on the degree of the correlation, the oscillators could be synchronized in different forms, such as complete synchronization, phase synchronization, and generalized synchronization~\cite{SYNREV:Boccaletti}. Among them, complete synchronization has the strongest correlation, as in complete synchronization the states of the oscillators are identical during the time course of system evolution~\cite{SYNREV:Pecora}. Recently, stimulated by the discoveries of the small-world and scale-free features in many realistic systems, synchronization behaviors in complex networks have received considerable attention~\cite{SYNREV:Arenas}. 

In the study of complete synchronization, a typical model employed in literature is an ensemble of identical chaotic oscillators coupled through a linear function (i.e., the diffusive coupling), and one of the central tasks is to predict, theoretically or numerically, the critical coupling strength for generating synchronization~\cite{Syn:Book}. When the dynamical equations of the system are available, the critical coupling in many cases can be predicted theoretically, e.g., by the method of master-stability function~\cite{MSF:Pecora,MSF:HL}. Yet in realistic situations the dynamical equations are mostly unknown, which means that the prediction should be made based on solely the time series measured from the system evolutions. One approach to address this question is to reconstruct first the system dynamics by techniques such as compressive sensing, and then estimate the critical coupling by model simulations~\cite{SRQ:2012}. A different approach proposed recently to address this question is exploiting RC, in which the synchronization error between the oscillators under arbitrary coupling strength can be calculated directly from the machine outputs~\cite{RC:FHW2021}. In these studies, a common requirement is that the time series must be measured from the desynchronization states. This requirement, however, is not met in many physical and biological systems, e.g., the power-grids and the circadian clock, whose normal functions rely on the synchronized motion of the dynamical units. Assuming that these systems are already operating on the stable synchronous states, i.e., the systems will restore to the synchronous states quickly after perturbations, the question we ask here is: given the transient time series and the coupling parameter of several states in the synchronization regime, can we predict whether the system will be desynchronized if a small drift is introduced to the coupling parameter? We are going to demonstrate that for the typical systems of coupled chaotic oscillators, the prediction can be accomplished by the scheme of parameter-aware RC introduced in Sec. II.

\begin{figure}[tbp]
\begin{center}
\includegraphics[width=0.95\linewidth]{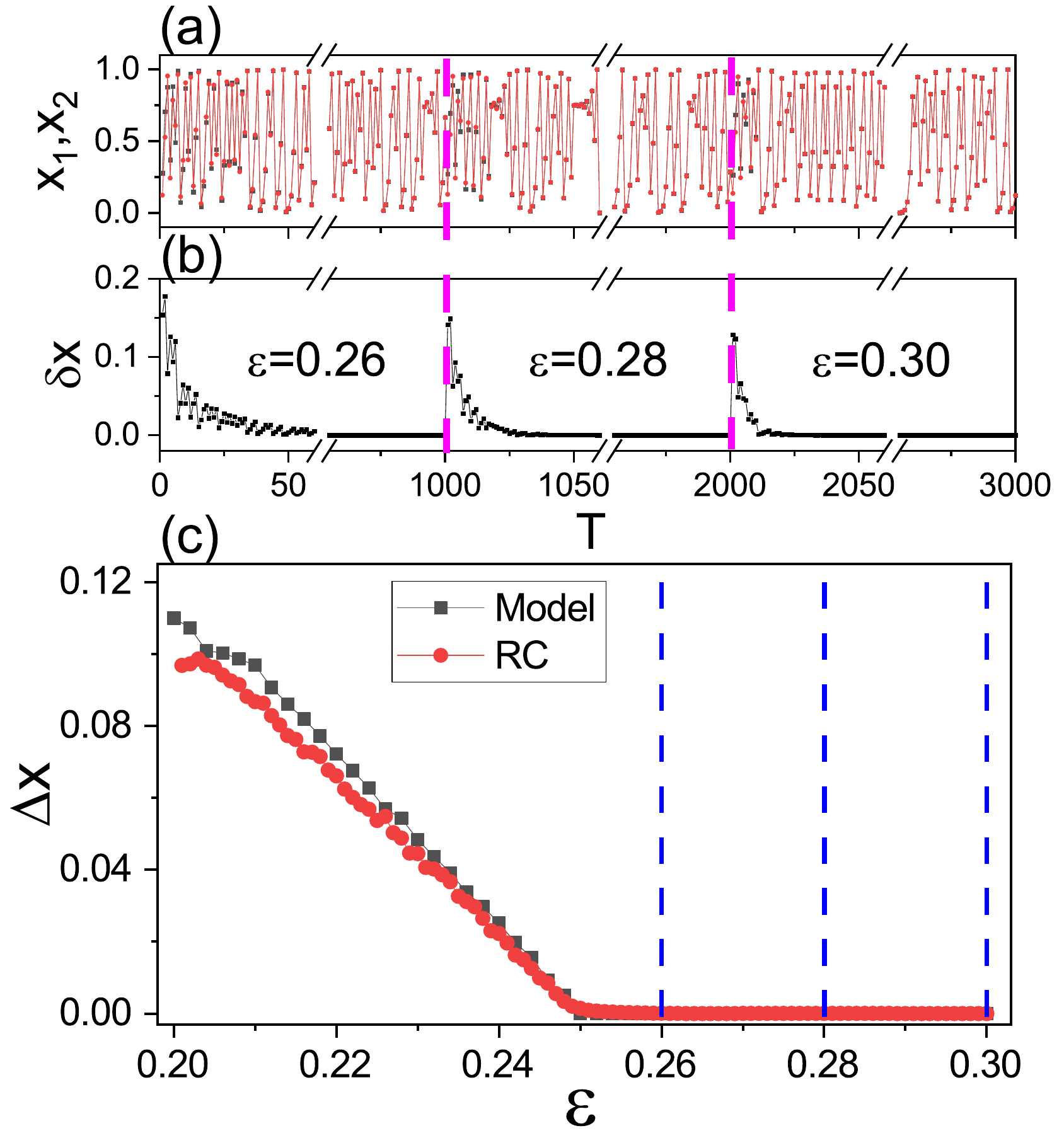}
\caption{Predicting synchronization transition in two coupled chaotic Logistic maps. (a) The training time series, which is generated by combining the time series of three sampling states in the synchronization regime, $\varepsilon=0.26$, $0.28$ and $0.30$. (b) The time evolution of the synchronization error $\delta x=|x_1-x_2|$ for the sampling states. (c) The variation of the time-averaged synchronization error $\Delta x$ with respect to $\varepsilon$ around the critical point $\varepsilon_1=0.25$. Black squares are the results obtained by model simulations. Red discs are the results predicted by the machine. Vertical dashed lines denote the locations of the sampling states.}
\label{fig4}
\end{center}
\end{figure}

We consider first the model of two coupled chaotic Logistic maps. The system dynamics reads
\begin{equation}\label{coupledlog}
x(n+1)_{1,2}=F[x(n)_{1,2}]+\varepsilon[H(x(n)_{2,1})-H(x(n)_{1,2})],
\end{equation}
where $x(n)$ denotes the state of the map at the $n$th iteration, $F(x)=4x(1-x)$ is the mapping function describing the local dynamics, $H(x)=F(x)$ is the coupling function, and $\varepsilon$ is the coupling strength. The initial states of the maps are chosen randomly from the interval $(0,1)$. When the system equations are available, the critical coupling for synchronization can be analyzed by the method of master stability function~\cite{MSF:Pecora,MSF:HL}, which shows that complete synchronization occurs in a bounded region in the parameter space, $\varepsilon \in (\varepsilon_{1}, \varepsilon_{2})$, with $\varepsilon_{1}=0.25$ and $\varepsilon_{2}=0.75$. To generate the training data from the synchronization regime, we choose $\varepsilon=0.26$, $0.28$ and $0.30$ as the sampling parameters. For each sampling parameter, we record the system state, $\mathbf{u}(t)=[x_1,x_2]^{T}$, for a period of $T=L'=1\times 10^3$ iterations. The combination of the three time series forms the training series, which is plotted in Fig.~\ref{fig4}(a). Figure~\ref{fig4}(b) plots the time evolution of the synchronization error, $\delta x=|x_1-x_2|$, for different parameters. We see that in all three cases the value of $\delta x$ is decreased to $0$ after a short transient ($T<50$). Figure~\ref{fig4}(b) shows also that with the increase of $\varepsilon$, the lifetime of the transient is shortened. The shortened transient at larger coupling strengths is contributed to the decreased conditional Lyapunov exponent. Specifically, as $\varepsilon$ increases from $\varepsilon_c$ in the synchronization regime, the Lyapunov exponent characterizing the stability of the dynamics in the phase space transverse to the synchronous manifold will be decreased from $0$ gradually, while the more negative is the exponent, the quicker will be the system returning to the synchronous state and the shorter will be the transient lifetime.   

The training series, together with the corresponding time series of the coupling strength $\varepsilon(n)$, are then fed into the machine for obtaining the output matrix. In this application, the optimized hyperparameters of the RC are $(D_{r}, p, \sigma, \lambda, \eta)=(100, 0.1, 1, 1\times 10^{-5}, 1\times 10^{-7})$. In obtaining the output matrix, a short segment of $\tau=5$ data points in each sampling series is used to drive the reservoir out of the transient states. In the predicting phase, we set $\varepsilon$ as the control parameter and tune it from $0.30$ to $0.20$ by the step $\Delta \varepsilon=2\times 10^{-3}$. For each value of $\varepsilon$, we first run the machine for a transient period of $T'_0=1\times 10^3$ iterations in order of removing the impact induced by the initial conditions, and then calculate from the machine outputs the time-averaged synchronization error $\Delta x=\langle|x_{1}-x_{2}|\rangle_T$, with $T=1\times 10^4$. The variation of $\Delta x$ with respect to $\varepsilon$ is plotted in Fig.~\ref{fig4}(d), which shows that the two maps are desynchronized at the critical coupling $\varepsilon_c\approx 0.25$ and, as $\varepsilon$ decreases from $\varepsilon_c$, the value of $\Delta x$ is gradually increased. To evaluate the prediction performance, we plot in Fig.~\ref{fig4}(d) also the results obtained from model simulations. We see that the predicted results are in good agreement with the results obtained from model simulations.

We consider next the model of two coupled chaotic Lorenz oscillators. The system dynamics reads
\begin{align}\label{lorenz}
&\dot{x}_{1,2}=\alpha(y_{1,2}-x_{1,2})+\varepsilon(x_{2,1}-x_{1,2}) \notag, \\
&\dot{y}_{1,2}=x_{1,2}(\beta-z_{1,2})-y_{1,2}+\varepsilon(y_{2,1}-y_{1,2}), \\
&\dot{z}_{1,2}=x_{1,2}y_{1,2}-\gamma z_{1,2}+\varepsilon(z_{2,1}-z_{1,2})\notag.
\end{align}
The parameters of the oscillators are $(\alpha, \beta, \gamma)=(10, 28, 2)$, by which the motion of isolated oscillator is chaotic. Analysis based on the method of master stability function shows that the two oscillators are synchronized when $\varepsilon>\varepsilon_{c}\approx0.42$~\cite{MSF:Pecora,MSF:HL}. Still, our mission here is to predict the critical coupling $\varepsilon_{c}$ based on the time series acquired at several states in the synchronization regime.

\begin{figure}[tbp]
\begin{center}
\includegraphics[width=0.95\linewidth]{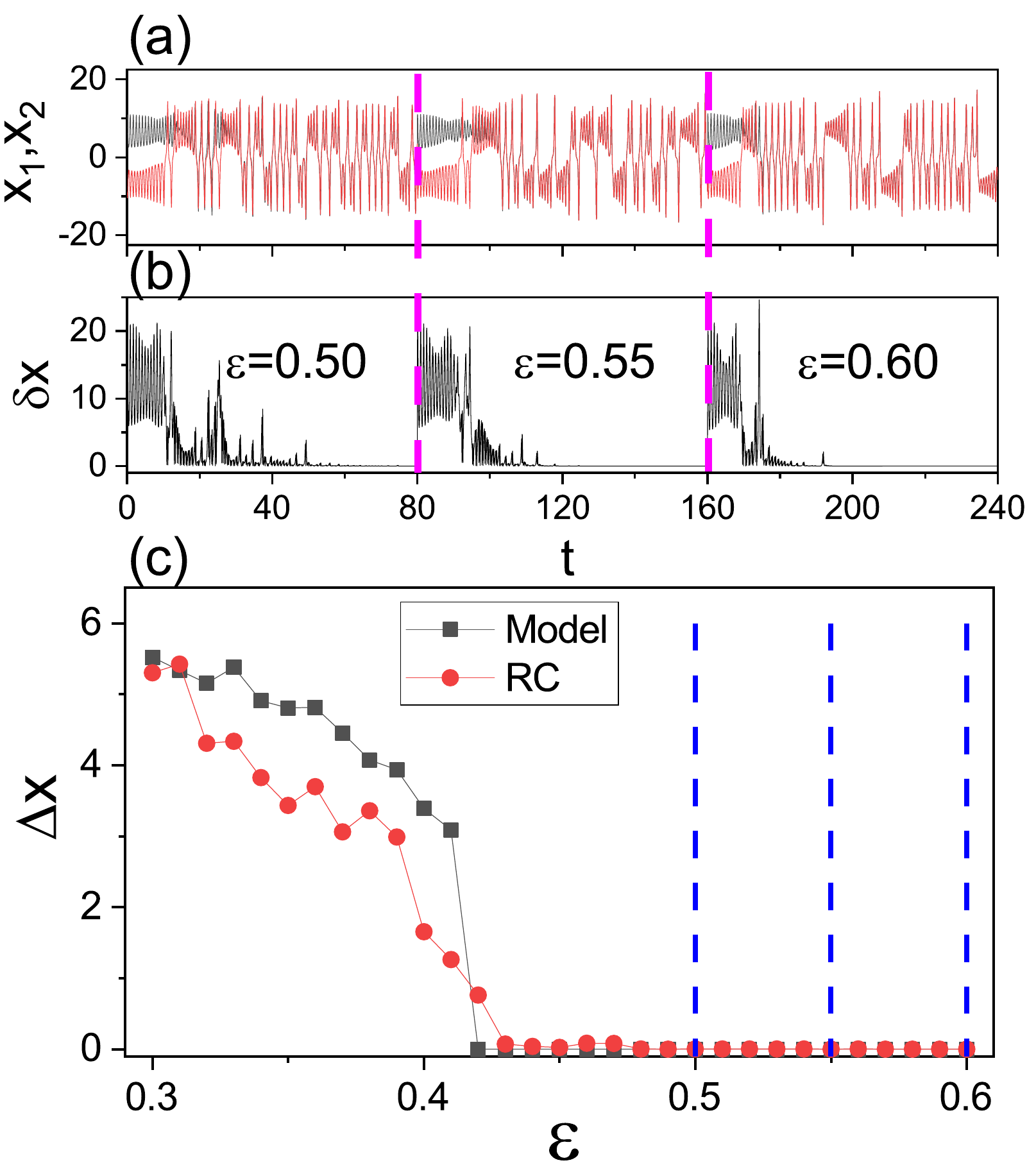}
\caption{Predicting synchronization transition in two coupled chaotic Lorenz oscillators. (a) The training series of the variables $x_1$ and $x_2$, which is generated by combining the time series of three sampling states in the synchronization regime, $\varepsilon=0.5$, $0.55$ and $0.6$. (b) The time evolution of the synchronization error $\delta x=|x_1-x_2|$ for the sampling states. (c) The variation of the time-averaged synchronization error $\Delta x$ with respect to $\varepsilon$ around the critical point $\varepsilon_c\approx 0.42$. Black squares are the results obtained from model simulations. Red discs are the results predicted by the machine. Vertical dashed lines denote the sampling parameters.}
\label{fig5}
\end{center}
\end{figure}

As the illustration, we choose $\varepsilon=0.5$, $0.55$ and $0.6$ as the sampling parameters (all within the synchronization regime). In model simulations, the initial conditions of the oscillators are randomly chosen within the range $(-1,1)$,  and the system is evolved according to Eq.~(\ref{lorenz}) by the time step $\Delta t=0.02$. For each sampling parameter, we record the system state, $\bm{u}(t)=[x_{1},y_{1},z_{1},x_{2},y_{2},z_{2}]^ {T}$, for a period of $T=80$. The combination of the recorded time series forms the training series, as depicted in Fig.~\ref{fig5}(a) for the variables $x_1$ and $x_2$. The time evolutions of the synchronization error $\delta x=|x_1-x_2|$ under different parameters are plotted in Fig.~\ref{fig5}(b). We see that in each case the synchronization error is damped to $0$ after a short transient ($T<60$). The training series and the corresponding time series of the coupling strength are then fed into the machine for estimating the output matrix. Here, a short segment of $\tau=50$ data points at the beginning of each sampling series is used to drive the reservoir out of the transient states. In this application, the hyperparameters of the RC are $(D_{r}, p, \sigma, \lambda, \eta)=(2\times 10^3, 0.2, 0.05, 0.1, 1\times 10^{-7})$.

To predict the critical coupling by the scheme of parameter-aware RC, we set $\varepsilon$ as the control parameter and decrease it from $0.6$ to $0.3$ by the step $\Delta \varepsilon=0.1$. For each value of $\varepsilon$, we first run the machine for a transient period of $T'_0=1\times 10^3$, and then calculate form the machine outputs the time averaged synchronization error $\Delta x= \langle|x_{1}-x_{2}|\rangle_{T}$, with $T=1\times 10^{3}$. The variation of $\Delta x$ with respect of $\varepsilon$ is plotted in Fig.~\ref{fig4}(c), which shows that the system is desynchronized at the critical coupling $\varepsilon_c\approx 0.43$. To check the performance of the predictions, we plot in Fig.~\ref{fig4}(c) also the results obtained from model simulations. We see that the predicted results agree with the results of model simulations reasonably well. 
 
For the case of synchronization transition, additional simulations have been also conducted to check the impacts of the sampling states on the prediction performance (not shown). It is found that for the fixed number of sampling states, the prediction performance becomes worse when choosing the larger sampling parameters. In particular, it is found that for the model of coupled Logistic maps, the machine fails to predict the second transition point $\varepsilon_2\approx 0.75$, despite the number and location of the sampling parameters. A close look to the transient behaviors of the system nearby $\varepsilon_2$ shows that the failure is due to the ultrashort transient in approaching synchronization. In specific, starting from the random initial conditions, the synchronization error between two maps is decreased to $0$ after just several iterations ($\sim 5$). For such a short transient, the machine is not able to get from the training data sufficient information about the system dynamics, and therefore is not able to predict the transition. This finding is consistent with the findings in quorum-sensing transition.

\section{discussion and conclusion}\label{dc}

In exploiting RC for predicting chaotic systems, two general requirements are that (1) the measured time series should reflect the actual dynamics of the target system and (2) the measured time series should be sufficiently long. For complex systems of coupled oscillators, when the system dynamics is degenerated to a low-dimensional manifold in the phase space, e.g., a fixed point (the cases of quorum sensing and amplitude death) or the synchronous state (the case of complete synchronization), many information of the system dynamics are missed in the asymptotic dynamics. For instance, when two oscillators are quenched to the origin, we get no information about the system dynamics from the constant time series. This is one of the reasons why in previous studies of phase-transition prediction the training data are all collected from the oscillatory regime of high-dimensional dynamics~\cite{KLW:2021,RC:FHW2021,RC:RXiao2021}. Different from the previous studies, in the current study we predict the transition in the opposite direction, e.g., predicting the transition by the data collected from the regime of degenerated, low-dimensional dynamics. The key lies in the transient activities preceding the asymptotic dynamics. We have demonstrated in different models that the transient activities, although of short lifetimes, reflect the actual dynamics of the target system and, by the scheme of parameter-aware RC, can be exploited to predict the critical point of phase transition. In systems of coupled limit cycles, we have shown that by the transient activities acquired in the quiescent regime in which the asymptotic dynamics is represented by a fixed point, the trained machine is able to predict not only the transition point, but also the system dynamics in the oscillatory regime (e.g., the variation of the order parameters); in system of coupled chaotic oscillators, we have shown that the transient dynamics in the synchronization regime can be exploited to predict not only the critical point for synchronization, but also the behavior of the synchronization error in the desynchronization regime. In both cases, it is found that the prediction performance is strongly affected by the transients, i.e., by increasing the lifetime of the transients, the prediction performance will be gradually improved. Consider the facts that many real-world systems are operating at stable, low-dimensional states and transient activities are ubiquitously observed in these systems, the findings may have broad applications, e.g., predicting the tipping points of climate systems~\cite{Rev:EarthSystem2021} and identifying the complex structure of ecological networks~\cite{SYNREV:Arenas}.        

While our studies show preliminarily the feasibility of predicting phase transition based on the time series of transients, many questions remain open. One is about the generality of the findings. For illustrative purposes, we have employed the model of coupled oscillators to demonstrate the predictions, it is not clear whether the similar results can be found in other type of models, e.g., systems showing the phenomenon of supertransient~\cite{Suppertransient:1983}. Another question is about the connection between the transient lifetime and the prediction performance. Our studies show that to predict the transition point accurately, the sampling parameters should be close to the critical point. Normally, as the sampling parameter leaves away from the critical point, the lifetime of the transient will be decreased by an algebraic scaling law~\cite{Book:Lai}. When the sampling parameter is far away from the critical point, the lifetime of the transient will be too short to train the machine. This phenomenon has been observed in  predicting the second point $\varepsilon_2$ of synchronization transition in coupled chaotic Logistic maps. It remains as a challenge to predict the transition when the sampling states are not close to the critical point. It is our hope that these questions could be addressed by further studies.

To summarize, exploiting the technique of RC in machine leaning, we have shown that the critical point of phase transition in coupled oscillators can be predicted by the time series of transient activities in the stable regime. The predictions are demonstrated in three different models, including an ensemble of indirectly coupled periodic oscillators showing quorum-sensing transition, two coupled non-identical oscillators showing amplitude-death transition, and two coupled chaotic oscillators showing complete-synchronization transition. In all cases, it is found that the machine trained by the transient dynamics at several states in the stable regime is able to predict not only the transient evolution in the stable regime, but also the critical point where the system becomes unstable. Our study highlights the importance of transient dynamics in machine learning, and provides an alternative approach for the model-free prediction of phase transition in complex dynamical system.

This work was supported by the National Natural Science Foundation of China under the Grant Nos.~11875182 and 12105165.

\end{document}